\documentclass[twocolumn,prl,secnumarabic,
amsmath,amssymb,superscriptaddress,nofootinbib]
{revtex4-2}

\usepackage{graphicx} 
\usepackage{xcolor}
\usepackage{empheq}
\usepackage[compat=1.1.0]{tikz-feynman}
\newcommand{\rf}[1]{Fig.\,\ref{#1}}

\newcommand{\nn}{\nonumber\\}
\newcommand{\ul}{\underline}

\newcommand{\bea}{\begin{eqnarray}}
\newcommand{\ea}{\end{eqnarray}}
\newcommand{\eea}{\end{eqnarray}}
\newcommand{\ord}{\,{\cal O}}

\begin{document}

\title{Search for axion resonances in vacuum birefringence with three-beam collisions}

\author{Stefan Evans}
\email{s.evans@hzdr.de}
\affiliation{Helmholtz-Zentrum Dresden-Rossendorf, Bautzner Landstraße 400, 01328 Dresden, Germany}
\author{Ralf Sch\"utzhold}
\affiliation{Helmholtz-Zentrum Dresden-Rossendorf, Bautzner Landstraße 400, 01328 Dresden, Germany}
\affiliation{Institut f\"ur Theoretische Physik, Technische Universit\"at Dresden, 01062 Dresden, Germany}

\begin{abstract}
We consider birefringent (i.e., polarization changing) scattering of x-ray 
photons at the superposition of two optical laser beams of ultra-high intensity 
and study the resonant contributions of axions or axion-like particles, 
which could also be short-lived. 
Applying the specifications of the Helmholtz International Beamline for 
Extreme Fields (HIBEF), we find that this set-up can be more sensitive 
than previous light-by-light scattering (birefringence) or light-shining-through-wall 
experiments in a certain domain of parameter space. 
By changing the pump and probe laser orientations and frequencies, 
one can even scan different axion masses. 
\end{abstract}

\date{\today}

\maketitle

\paragraph{Introduction}

After the discovery of the Higgs particle~\cite{Higgs:1964pj}, 
axions or axion-like particles 
are one of the most favorite candidates for new physics beyond the standard 
model.  
One way to motivate them is to consider the electromagnetic field strength 
tensor $F_{\mu\nu}$ and its dual $\tilde F_{\mu\nu}$ which can be contracted 
to yield the two lowest-order Lorentz invariants 
$F_{\mu\nu}F^{\mu\nu}=2({\bf B}^2-{\bf E}^2)
=-\tilde F_{\mu\nu}\tilde F^{\mu\nu}$ as well as 
$\tilde F_{\mu\nu}F^{\mu\nu}=-4{\bf B}\cdot{\bf E}$. 
The former generates the Lagrangian density of electromagnetism while 
the latter is usually discarded because it is a total derivative.
However, this argument is only valid if the pre-factor in front of this term 
$\tilde F_{\mu\nu}F^{\mu\nu}$ is a constant. 
If this pre-factor is a dynamical field $\phi$, i.e., space-time dependent, 
this term does generate a non-trivial (effective) interaction Lagrangian 
of the form ($\hbar=c=1$)
\bea
\label{Lagrangian}
{\cal L}_{\rm int}=g_\phi\phi\,{\bf B}\cdot{\bf E}
\,,
\ea
where $g_\phi$ denotes the (effective) interaction strength. 
Since the term $\tilde F_{\mu\nu}F^{\mu\nu}$ is odd under parity ${\cal P}$,
the axion field $\phi$ is usually considered a pseudoscalar field.

Apart from this effective field theory approach, axions were originally 
proposed as a possible solution to the strong ${\cal CP}$ problem in 
quantum chromo-dynamics (QCD), see, 
e.g.,~\cite{Peccei:1977hh, Weinberg:1977ma, Wilczek:1977pj, Kim:1979if, 
Shifman:1979if, Zhitnitsky:1980tq, Dine:1981rt}.
In the following, axions and axion-like particles will be used synonymously.
Modeling the axion field as a massive scalar field weakly coupled to the 
other standard model particles, it could also be a candidate for dark 
matter~\cite{Preskill:1982cy, Abbott:1982af, Dine:1982ah, Sikivie:1983ip, 
Raffelt:1996wa, Raffelt:2006cw, Ringwald:2012hr, Ouellet:2018beu, 
Caputo:2018vmy, Alonso-Alvarez:2019ssa, Carenza:2019vzg, HAYSTAC:2020kwv, 
Semertzidis:2021rxs} 
and would have important consequences for cosmology, see,
e.g.,~\cite{Jaeckel:2010ni, Baker:2013zta, Irastorza:2018dyq, 
Buschmann:2019icd, Cadamuro:2011fd, DeAngelis:2007dqd, 
Fortin:2021cog}.

In search of observable effects, astronomical data provide very important 
sources~\cite{Bernabei:2001ny, Ayala:2014pea, ParticleDataGroup:2022pth}. 
Similar to neutrinos, weakly coupled and long lived axions could provide 
a cooling mechanism for stars and other astrophysical objects (such as 
white dwarfs~\cite{Isern:2008nt}), mostly due to their 
coupling~\eqref{Lagrangian} to photons. 
In fact, the apparent absence of such effects for our sun, for example,  
leads to significant restrictions on the parameter space of 
axions~\cite{CAST:2017uph}, see also~\cite{XENON:2020rca, Dent:2020jhf}.  

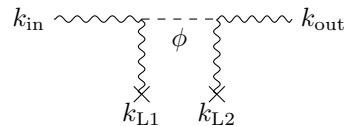
\begin{figure}[h]
\begin{tikzpicture}
\begin{feynman}
\vertex (a) {\normalsize\(k_{{\rm in}}\)};
\vertex [right=1.5cm of a] (b); 
\vertex [below=1cm of b] (f1) {\normalsize\(k_{{\rm L}1}\)}; 
\vertex [right=1cm of b] (c);
\vertex [right=1cm of c] (f2) {\normalsize\(k_{{\rm out}}\)};
\vertex [below=1cm of c] (f3) {\normalsize\(k_{{\rm L}2}\)};
\diagram* {
(a) -- [photon] (b),
(b) -- [photon, insertion={[line width=1mm]1}] (f1), 
(b) -- [scalar, edge label'=\normalsize\(\phi\)] (c), 
(c) -- [photon] (f2),
(c) -- [photon, insertion={[line width=1mm]1}] (f3),
}; 
\end{feynman}
\end{tikzpicture} 
\caption{\label{AmplitudeFeynman}{Axion $s$-channel contribution to 
light-by-light scattering. 
The initial and final x-ray photons with momenta $k_{{\rm in}}$ and 
$k_{{\rm out}}$ interact with the fields of the two optical lasers 
$k_{{\rm L}1}$, $k_{{\rm L}2}$ via the internal axion propagator 
(dashed line).}} 
\end{figure}

Nevertheless, such astronomical observations cannot supersede laboratory 
experiments. On the one hand, a direct and active experimental manipulation is 
qualitatively different from an indirect observation, in particular 
since our conclusions drawn from the latter depend on our correct 
understanding of stellar dynamics etc. 
On the other hand, there are many reasons why axions detected in the 
laboratory could still be consistent with astronomical observations 
(especially if they occur on very different scales)
\cite{Ahlers:2006iz, Brax:2007vm, Brax:2007ak, Dupays:2006dp, 
Gies:2006ca, Jaeckel:2006xm, Liao:2007nu, Mohapatra:2006pv}, for example 
interaction effects such as running coupling constants or axion 
confinement~\cite{Jain:2005nh, Masso:2005ym, Masso:2006gc}.

The question of axion lifetimes and length scales distinguishes two 
major classes of laboratory-based experiments. 
Akin to astronomical searches, one class looks for long-lived axions 
propagating over macroscopic distances, including ``light shining through wall'' 
experiments~\cite{Ruoso:1992ai, Cameron:1993mr, Fouche:2008jk, NOMAD:2000usb, 
OSQAR:2007oyv, Redondo:2010dp, GammeVT-969:2007pci, Afanasev:2008jt, 
Ehret:2010mh, OSQAR:2013jqp, OSQAR:2015qdv, Inada:2016jzh, Yamaji:2018ufo, 
Beyer:2021mzq, Ortiz:2020tgs} 
based on the creation of axions from electromagnetic fields via the 
coupling~\eqref{Lagrangian}. 
Then, after propagating through the wall, the axion is converted back into an electromagnetic signal. 
A related but more indirect mechanism is based on detecting 
``missing photon energy'', e.g., 
at the BABAR experiment~\cite{Dolan:2017osp}. Photons produced in electron-positron collisions could undergo axion  Bremsstrahlung~\cite{Tsai:1986tx}. 
The signature of the generated axions escaping the  detector would 
then be an observable energy loss.

As a qualitatively different class of scenarios, sensitive to much 
smaller length scales, effective 
photon-photon interactions (light-by-light scattering) could be mediated 
by an internal 
axion line~\cite{Maiani:1986md, Raffelt:1987im, Semertzidis:1990qc, 
Bernard:1997kj, 
Villalba-Chavez:2013bda, Villalba-Chavez:2016hxw, Tommasini:2009nh, 
Dobrich:2010hi, Evans:2018qwy, 
Bogorad:2019pbu, Shakeri:2020sin,  Beyer:2021xql, SAPPHIRES:2021vkz, 
Ishibashi:2023bae}, see also~\rf{AmplitudeFeynman}. 
In this case, the axion does not propagate a macroscopic distance and thus 
such experiments would also be sensitive to axions which are not quasi-free 
and long-lived (at the scales relevant to the experiment).

Prominent examples for the second class are 
PVLAS~\cite{PVLAS:2005sku, PVLAS:2007wzd, Ejlli:2020yhk},  
BMV~\cite{Battesti:2008, Cadene:2013bva, Battesti:2018bgc} and 
OVAL~\cite{Fan:2017fnd}.
Using a strong and static magnetic field as the pump field for polarizing 
the vacuum, the goal was to detect this change with an optical laser 
as the probe field. 
The sought-after signal was then a rotation or flip of the optical laser 
polarization, i.e., quantum vacuum birefringence.

In this work, we study an alternative scenario which is motivated by a 
recent proposal~\cite{Ahmadiniaz:2022nrv} for detecting quantum vacuum 
birefringence as 
predicted by quantum electrodynamics (QED). 
As the probe field, we envision x-ray photons generated by an x-ray free 
electron laser (XFEL) because their high frequency increases the signal. 
The pump field is supposed to be a superposition of 
two optical lasers, which offer pump field strengths much larger than in PVLAS. 
As already proposed in~\cite{Ahmadiniaz:2022nrv}, 
the momentum exchange between the XFEL and the pump lasers facilitates a 
finite scattering angle (in the mrad regime) which helps us to discriminate 
the signal photons from the background (the main XFEL beam). 

\paragraph{Geometry}

To illustrate our main idea, let us start with the most simple set-up.
The initial x-ray photon is described by its energy $\omega_{\rm in}$, 
momentum ${\bf k}_{\rm in}=\omega_{\rm in}{\bf n}_{\rm in}$,
polarization ${\bf e}_{\rm in}$, and analogously for the final 
x-ray photon with $\omega_{\rm out}$, 
${\bf k}_{\rm out}=\omega_{\rm out}{\bf n}_{\rm out}$
and ${\bf e}_{\rm out}$,
as well as for the two optical lasers with the same frequency 
$\omega_{\rm L1}=\omega_{\rm L2}=\omega_{\rm L}$, 
but different propagation directions 
${\bf k}_{\rm L1,2}=\omega_{\rm L}{\bf n}_{\rm L1,2}$
and polarizations ${\bf e}_{\rm L1,2}$. 

In order to obtain a resonant enhancement of our signal (see below), 
while also maximizing the deflection angle of the signal XFEL photon, 
we consider the case where an optical photon is absorbed from one 
laser and emitted into the other, such that energy and momentum 
conservation read 
\bea
\label{conservation}
&& 
\omega_{\rm out}
=
\omega_{\rm in}+\omega_{\rm L1}-\omega_{\rm L2}
=
\omega_{\rm in} 
\;,
\nn
&& {\bf k}_{\rm out}={\bf k}_{\rm in}+{\bf k}_{\rm L1}-{\bf k}_{\rm L2} 
\,.
\ea
Since we focus on the dominant (resonant) axion contribution and 
the birefringent ${\bf e}_{\rm in}\perp{\bf e}_{\rm out}$ signal 
in or close to forward direction ${\bf n}_{\rm in}\approx{\bf n}_{\rm out}$, 
the direct interaction~\eqref{Lagrangian} between the initial and final
x-ray photons is suppressed and hence 
we focus on the $s$-channel in Fig.~\ref{AmplitudeFeynman} and neglect the 
$t$-channel.

As a consequence, each vertex~\eqref{Lagrangian} combines an x-ray 
photon with either of the two optical lasers. 
By adjusting the polarization and propagation unit vectors appropriately, 
we can select the various possibilities. 

\begin{figure}[h]
\center
\includegraphics[width=.95\columnwidth]{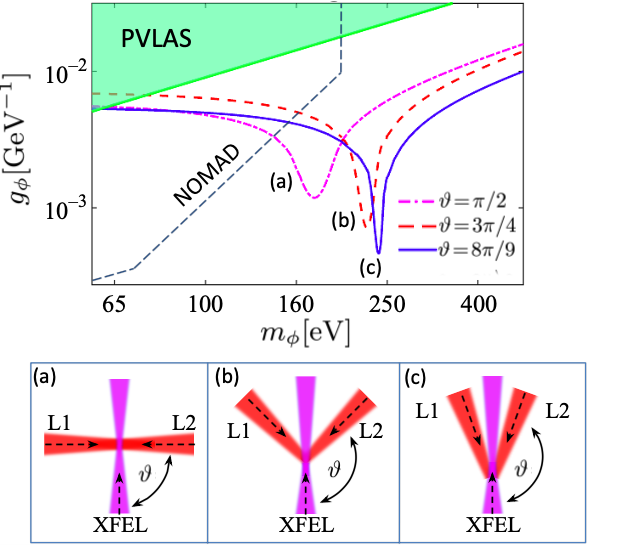}
\caption{\label{parameterspace}{
Accessible parameter space based on $N_{\rm signal}\geq1$ from 
Eq.~\eqref{photonProb} in terms of axion mass $m_\phi$ and coupling $g_\phi$. 
The optical laser orientations relative to the XFEL (at $\vartheta=0$) are 
$\vartheta=8\pi/9$ (blue solid curve), 
$\vartheta=3\pi/4$ (red dashed curve) and 
$\vartheta=\pi/2$ (purple dot-dashed curve). 
The green shaded region in the top left corner denotes the parameter 
region probed by PVLAS (birefringence~\cite{Ejlli:2020yhk}).
The limits obtained by NOMAD (light-shining-through-wall \cite{NOMAD:2000usb}) are given by the black dashed curve. 
}}
\end{figure}

Let us first consider the fully perpendicular crossed-beam case 
${\bf n}_{\rm L1,2}\perp {\bf n}_{\rm in}$ 
where the two optical lasers collide head-on 
${\bf n}_{\rm L1}=-{\bf n}_{\rm L2}$, see Fig.~\ref{parameterspace}a. 
If we choose ${\bf e}_{\rm L1}={\bf e}_{\rm L2}={\bf e}_{\rm in}$,
there would be no axion contribution at all. 
Rotating the optical laser polarizations to 
${\bf e}_{\rm L1}={\bf e}_{\rm L2}={\bf n}_{\rm in}$, 
while keeping ${\bf e}_{\rm in}$ fixed, 
the axion interaction~\eqref{Lagrangian} would lead to 
scattering with the same 
polarization ${\bf e}_{\rm out}\|{\bf e}_{\rm in}$.
A birefringent signal ${\bf e}_{\rm out}\perp{\bf e}_{\rm in}$
could be obtained after rotating ${\bf e}_{\rm in}$ by $45^\circ$
for example. 

However, if we tilt the optical laser beams more towards the axis 
which is counter-propagating to the XFEL, as in Fig.~\ref{parameterspace}c, 
the birefringent signal ${\bf e}_{\rm out}\perp{\bf e}_{\rm in}$ would 
actually dominate for crossed optical laser polarizations 
${\bf e}_{\rm L1}\perp{\bf e}_{\rm L2}$ since then 
${\bf e}_{\rm out}$ and ${\bf e}_{\rm in}$ could be aligned with
either ${\bf e}_{\rm L1}$ or ${\bf e}_{\rm L2}$, respectively.

\paragraph{Axion Propagator}

The lowest-order Feynman diagram of the process under consideration is 
displayed in~\rf{AmplitudeFeynman}. 
In terms of the momentum four-vectors $\ul{k}_{\rm in}$ and $\ul{k}_{\rm L 1}$, 
the four-momentum of the internal axion line reads 
$\ul{p}_\phi=\ul{k}_{\rm in}+\ul{k}_{\rm L1}$ and thus its contribution 
to the amplitude becomes 
\bea
\label{amplitude}
\frac{g_\phi^2}{(\ul{k}_{\rm in}+\ul{k}_{\rm L1})^2-m_\phi^2}
=
\frac{g_\phi^2}{
{2(}\omega_{\rm in}\omega_{\rm L}-{\bf k}_{\rm in}\cdot{{\bf k}_{\rm L1}})-m_\phi^2}
\,,
\ea
where we have assumed that axion can be described by the standard propagator 
of a scalar field with mass $m_\phi$. 

For the crossed-beam geometry discussed above, we have 
${\bf k}_{\rm in}\perp{\bf k}_{\rm L1}$ 
and thus the amplitude would be enhanced strongly near the resonance 
$2\omega_{\rm in}\omega_{\rm L}\approx m_\phi^2$ corresponding to an axion mass 
of order $\ord(10^2~\rm eV)$. 
By varying the angles between the optical lasers and the XFEL, one can scan 
different axion masses (see below). 

In fact, exactly on resonance $2\omega_{\rm in}\omega_{\rm L}=m_\phi^2$, 
the amplitude would actually diverge in the case of perfect plane waves. 
Of course, this implies that higher orders in $g_\phi$ should be taken 
into account.
A simple way of effectively doing this is to include self-energy terms 
in the propagator containing an imaginary part 
which then corresponds to a decay rate $\Gamma_\phi\sim g_\phi^2$. 
For plane waves, this would imply that the amplitude~\eqref{amplitude}
is highly sensitive to the value of $\Gamma$. 
However, the optical laser is not a perfect plane wave, but a focused beam 
-- with finite energy-momentum spread, 
which regularizes the amplitude~\eqref{amplitude}. 
This removes dependence on $\Gamma$ (unless it is larger than the 
energy-momentum spread of the optical laser)
and thus accommodates both long and short-lived axions.

\paragraph{Amplitude}

Combining the coupling~\eqref{Lagrangian} with the 
propagator~\eqref{amplitude}, the $s$-channel amplitude reads 
\bea
\label{amplitude-full}
{\mathfrak A}^{\rm s}
&=&
g_\phi^2
{ \dfrac{({\bf e}_{\rm in}\cdot[(\omega_{\rm in}{\bf k}_{\rm L1}-\omega_{\rm L1}{\bf k}_{\rm in})\times{\bf A}_{\rm L1}])}
{ 2(\omega_{\rm in}\omega_{\rm L}-{\bf k}_{\rm in}\cdot{\bf k}_{\rm L1})-m_\phi^2} }
\nn &&
{ \times
({\bf e}_{\rm out}\cdot[(\omega_{\rm out}
{\bf k}_{\rm L2}-\omega_{\rm L2}{\bf k}_{\rm out})\times{\bf A}_{\rm L2}]) }
\,,
\ea
where ${\bf A}_{\rm L1,2}$ denote the vector potentials of the optical lasers. 
As explained above, the realistic description of a laser focus which is 
localized in space requires the average over a finite momentum spread 
$\int d^3 k_{\rm L}$, which we implement 
with a distribution function ${\bf A}_{\rm L}({\bf k}_{\rm L})$. 
This averaging procedure then also regularizes the resonant singularity 
of the axion propagator at 
${2(\omega_{\rm in}\omega_{\rm L}-{\bf k}_{\rm in}\cdot{{\bf k}_{\rm L1}})}=m_\phi^2$.

A finite temporal duration of the optical laser pulse would correspond to a 
spread in frequencies $\omega_{\rm L}=|{\bf k}_{\rm L1,2}|$ 
but we neglect this spread here and focus on a fixed frequency 
$\omega_{\rm L}={|{\bf k}_{\rm L1,2}|}$ for simplicity. 

\paragraph{Experimental Parameters}

Taking the specifications of the Helmholtz International Beamline for 
Extreme Fields (HIBEF) as an example, we consider the following experimental 
setup, as illustrated in~\rf{FigSetup}.
The optical lasers are characterized by their frequency 
$\omega_{{\rm L}}=1.5~{\rm eV}$, focus intensity 
${\bf E}^2=4\times10^{21}{\rm W/cm^2}$, 
with a $3~\mu$m waist and a divergence of $\pm15$ degrees. 
We model the optical laser focus by a superposition of plane waves 
with the same frequency $\omega_{{\rm L}}$ and a Gaussian distribution 
for the transversal momentum spread. 
Assuming a repetition rate of 5~Hz~\cite{Zastrau:2021}, 
one could carry out an experiment with ${\cal O}(10^4)$  
shots in less than one hour, such that we set ${\cal O}(10^{-4})$ 
birefringent x-ray photons per shot as our detection threshold.

%
\begin{figure}[h]
\center
\includegraphics[width=0.95\columnwidth]{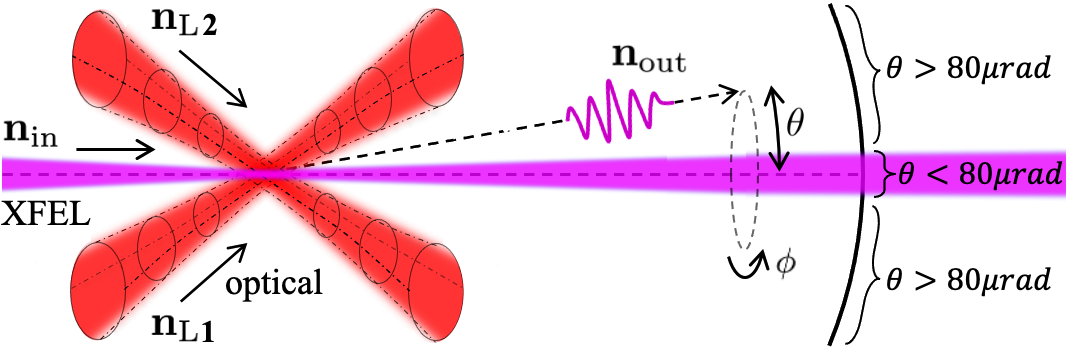}
\caption{\label{FigSetup}{
Sketch of the experimental set-up.
}} 
\end{figure}
%
%

We probe the optical laser background using an XFEL pulse of frequency 
$\omega_{{\rm in}}=10~{\rm keV}$, 
comprising $N_{{\rm XFEL}}=10^{12}$ photons per shot, 
with a beam waist of $5~\mu{\rm m}$ 
and a $80~\mu$rad beam divergence \cite{Ahmadiniaz:2022nrv, Zastrau:2021}.
The combined momentum transfer supplied by pump field, 
being in the optical regime, scatters the XFEL photons outside of this 
$80~\mu$rad cone. 
Combining this consideration with 
energy-momentum conservation which allows a maximum scattering angle of 
$300~\mu$rad, we thus search for a signal between 
$80~\mu{\rm rad}<\theta<300~\mu{\rm rad}$.  

\paragraph{Axion Signal}

Now we are in the position to estimate the signal strength. 
As motivated above, we focus on the $s$-channel amplitude as the dominant 
contribution. 
Although only the case of the first photon being 
absorbed and the second one emitted yields a resonant enhancement 
and is thus the most important contribution, we also include the opposite 
(emission first) case for the sake of completeness and sum over both cases. 
Furthermore, we sum the diagram in Fig.~\ref{AmplitudeFeynman} and the 
reverse sequence (exchanging the two optical lasers). 
Averaging the optical photons over the transverse momentum spread, 
we obtain the polarization-conserving (${\bf e}_{{\rm in}}\|{\bf e}_{{\rm out}}$) 
as well as birefringent (${\bf e}_{{\rm in}}\perp{\bf e}_{{\rm out}}$) 
differential cross sections as 
\bea
\label{ALPdiffcross}
\frac{d\sigma}{d\Omega}=
\sum_\pm \frac{|{\mathfrak A}^{\rm s}_\pm|^2}{4(2\pi)^3}
\,,
\ea
where we have used that $\omega_{\rm out}/\omega_{\rm in}\approx1$.
Subscripts $\pm$ label summation over both sequences of absorbed and 
emitted photons.

Given energy-momentum conservation, the XFEL can deflect to the left or right, 
e.g. parallel to the first (absorbed) optical photon and opposite the second 
(emitted), for the fully perpendicular case 
${\bf n}_{\rm L1,2}\perp {\bf n}_{\rm in}$ with 
${\bf n}_{\rm L1}=-{\bf n}_{\rm L2}$. 
By tuning the polarizations, i.e. choosing which of ${\bf e}_{\rm L1}$ or 
${\bf e}_{\rm L2}$ is aligned with ${\bf e}_{\rm out}$ or ${\bf e}_{\rm in}$, 
we determine the sequence in which the photons interact and thus which 
way the signal photons deflect. 

For pure plane waves, one could envision laser polarizations to be exactly 
aligned or orthogonal to the XFEL's, completely filtering the deflection 
in one direction. 
In our case however, the transverse momentum spread of the photons also 
means a distribution of their polarizations, so there is always some 
nonzero alignment with the XFEL polarization. 
Maximizing and minimizing the two laser alignments respectively allows 
us to tune the signal to prefer one direction by two orders of magnitude.

To determine the total number $N_{\rm signal}$ of signal photons, 
we integrate~\eqref{ALPdiffcross} over the domain of scattering angles 
discussed above $80~\mu{\rm rad}<\theta<300~\mu{\rm rad}$. 
Taking into account the XFEL photon number $N_{\rm XFEL}$ per shot 
multiplied by the number $10^4$ of shots and the size of the XFEL focus 
$w_{{\rm XFEL}}=5~\mu{\rm m}$, we find 
\begin{align}
\label{photonProb}
N_{\rm signal}\approx 
10^4\frac{N_{{\rm XFEL}}
}{w_{{\rm XFEL}}^2} 
\int\limits_0^{2\pi} d\phi 
\int\limits_{8\cdot10^{-5}}^{{3\cdot10^{-4}}}
d\theta\sin\theta\, 
\frac{d\sigma}{d\Omega}
\;.
\end{align}
As a function of $m_\phi$, the signal strength $N_{\rm signal}$ is peaked 
at resonance for a given laser geometry~\eqref{amplitude}. 
Furthermore, the on-shell requirement for the XFEL and optical photons, 
e.g.,  $|{\bf k}_{\rm L1}|=|{\bf k}_{\rm L2}|=\omega_{\rm L}$ 
in conjunction with energy-momentum conservation~\eqref{conservation} 
generates a substructure consisting of much narrower peaks within the 
resonance. 
However, since the optical laser pulses will inevitably display small 
variations during the $10^4$ shots, this small-scale substructure averages 
out -- which we model by a Gaussian convolution.

In Fig.~\ref{parameterspace} we plot the domain of accessible axion parameter 
space as the coupling $g_\phi$ and mass $m_\phi$ for which $N_{\rm signal}\geq1$ 
in~\eqref{photonProb}, i.e., one or more signal photons per $10^4$ XFEL shots. 
We display three optical laser orientations, 
the fully perpendicular case ${\bf k}_{\rm in}\perp{{\bf k}_{\rm L1,2}}$,
here labeled $\vartheta=\pi/2$, as well as the cases {$\vartheta=3\pi/4$} and {$\vartheta=8\pi/9$}, 
where $\vartheta$ denotes the angle between the optical laser and the XFEL. 

%
%
%

As already discussed after Eq.~\eqref{amplitude}, varying this angle effectively amounts to scanning different ranges of the axion mass. Indeed, when going from $\vartheta={\pi/2}$ to ${8\pi/9}$, the resonance shifts to higher axion masses and becomes more narrow. As a result, the enhancement of the signal at resonance increases. E.g., the case ${8\pi/9}$ produces the strongest signal and is most sensitive to axion masses around $m_\phi={240}$~eV.

\begin{figure}[h]
\center
\includegraphics[width=.95\columnwidth]{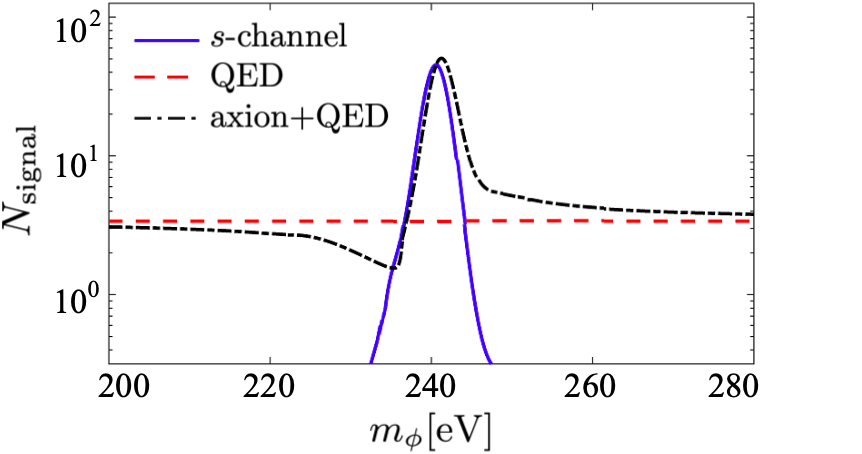}
\caption{\label{alldiagrams}
Signal strength $N_{\rm signal}$ from~\eqref{photonProb} for $\vartheta=8\pi/9$, plotted as a function of $m_\phi$ at fixed coupling $g_\phi=10^{-3}{\rm GeV}^{-1}$. 
The solid blue curve represents the axion $s$-channel contribution alone while the black dot-dashed curve incorporates the combined effect of axion $s$ and $t$-channels as well as the QED contribution (red dashed horizontal line).  
} 
\end{figure}

Far away from resonance, i.e., at lighter or heaver axion masses, QED birefringence becomes important. For the parameters used here, it can be estimated from the Euler-Heisenberg-Schwinger effective action~\cite{Heisenberg:1935qt, Weisskopf:1996bu, Schwinger:1951nm}, see also \cite{Ahmadiniaz:2022nrv}. Near resonance, combining the axion and QED Feynman diagram can also generate interference effects, see Fig.~\ref{alldiagrams}.

So far we have considered exclusively the birefringent signal -- 
one may also choose to include the polarization-conserving case as 
part of the desired signal. Applying e.g. polarizations 
${\bf e}_{\rm L1}={\bf e}_{\rm L2}={\bf n}_{\rm in}$, 
further enhancement of the signal strength is possible, 
since both lasers could then couple to the incoming XFEL, 
with their polarizations aligned to maximize the interaction. 
On the other hand, as discussed after Eq.~\eqref{conservation}, 
the choice ${\bf e}_{\rm L1}={\bf e}_{\rm L2}={\bf e}_{\rm in}$ 
suppresses the axion contribution, providing a diagnostic tool for filtering the pure QED signal.

%
%
%

\paragraph{Conclusions}

We have evaluated the axion contribution to birefringent
light-by-light scattering for an XFEL probe and optical laser pump. 
Special emphasis is placed on the resonant axion contribution 
which allows us to scan different axion masses by changing the 
involved parameters such as the angle $\vartheta$ between the 
XFEL and the optical laser. 
Furthermore, the axion resonance facilitates a sensitivity surpassing that of 
previous light-by-light scattering and light-shining-through-wall experiments such as PVLAS, BMV and NOMAD. 
Note that the parameters in Fig.~\ref{parameterspace}, including the 
bounds from PVLAS and NOMAD, lie inside the region tested by BABAR and 
solar axion searches such as CAST. 
However, as explained above, these are more indirect tests which rely on 
various assumptions such as long lived axions. 
Thus, the scheme presented here could provide the most stringent 
laboratory-based probe of short lived axion contributions.

Complementary to astrophysical bounds 
(e.g., \cite{ParticleDataGroup:2022pth}),  
such laboratory-based probes are also sensitive to axions which 
evade these bounds in some way. 
Examples could be interaction effects such as running coupling or 
confinement, see, e.g., \cite{Masso:2005ym, Masso:2006gc}, which 
invalidate the picture of long-lived and free-streaming axions. 
Although we treated the axion field as a free massive scalar 
field for simplicity, our results can be generalized to this case
by inserting the effective axion propagator into our amplitude.
If this propagator displays one or more quasi-particle peaks,  
we would again obtain axion resonances. 
The width of these quasi-particle peaks (related to their life-time) 
would then be added to the width generated by the angular spread 
of the optical laser. 

In view of the smallness of the signal, a discussion of its detectability 
should also include an estimate of possible background effects which might 
induce a false signal. 
These background effects are basically the same as already discussed 
in~\cite{Ahmadiniaz:2022nrv} devoted to the pure QED birefringence 
effect (see~\cite{Heisenberg:1935qt, Weisskopf:1996bu, Schwinger:1951nm, 
DiPiazza:2006pr, Heinzl:2006xc, Inada:2014srv, 
Schlenvoigt:2016jrd, Lundstrom:2005za, Inada:2017lop, DiPiazza:2007yx, 
King:2018wtn, Tommasini:2010fb, King:2012aw, Gies:2017ezf, Gies:2017ygp, 
Seino:2019wkb, 
Dobrich:2009kd, Gies:2014wsa, Grote:2014hja, Robertson:2020nnc, Gies:2022pla, 
Karbstein:2021ldz, Karbstein:2022uwf}). 
As discussed above, the axion signal displays distinctive dependence 
on the geometry (e.g., polarization directions), which could help to 
distinguish it from possible background effects.

In order to advance the sensitivity further, one could use more intense 
optical lasers (which will soon become available at HIBEF or at 
other facilities) or XFELs or tighten the XFEL beam waist~\cite{Zastrau:2021}, 
as well as increase the number of shots in the experiment.


\paragraph{Acknowledgments}

We thank 
N.~Ahmadiniaz, 
T.~Cowan, 
M.~Ding, 
S.~Franchino-Vi\~nas, 
H.~Gies,  
B.~King, 
J.~Grenzer, 
F.~Karbstein, 
M.~Lopez,
R.~Shaisultanov,
M.~\v{S}m\'id, 
and 
T.~Toncia
for fruitful discussions.  
This research was supported in part by Perimeter Institute for 
Theoretical Physics. 
Research at Perimeter Institute is supported by the Government 
of Canada through the Department of Innovation, 
Science and Economic Development and by the Province of Ontario 
through the Ministry of Research and Innovation.
R.S.~acknowledges funding by the Deutsche Forschungsgemeinschaft 
(DFG, German Research Foundation) -- Project-ID 278162697-- SFB 1242. 
%




\end{document}